\documentclass[twocolumn,twoside]{article}
\usepackage{graphicx}
\title{\Large \bf  Bivector Gauge Fields and Amplification of Radiation \\ from Distant Cosmic Sources}
\author{\normalsize Alexander Krasulin \\ \normalsize \it Institute for Nuclear Research of the
Russian Academy of Sciences\footnote{Former affiliation.} \\ \normalsize krasulin@post.com}
\date{\normalsize \bf Abstract \\ \mbox{ } \\ \begin{minipage}{400pt}
\normalsize \rm This paper describes in some detail the concept of bivector gauge fields and then examines the possibility of confirming their existence by recording certain patterns in variation of intensity of radiation from distant cosmic sources versus radiation frequency and versus the estimated distance to the source. \end{minipage} }
\begin{document}

\maketitle

This paper deals with the so-called {\it bivector} gauge fields; more precisely, with those of their kind that have to do with electromagnetism. The concept of bivector gauge fields arises rather naturally within the theory of five-dimensional tangent vectors in space-time, presented in ref.[1]\footnote[2]{A brief introduction can be found in ref.[2].}. However, it is quite possible to explain what these gauge fields are in terms of ordinary four-vectors and four-tensors as well. Though in this case it will not be so clear why they have been named `bivector'. Nor why anyone should have thought of a thing like that in the first place!

To explain what bivector gauge fields are one should recall what is parallel transport. As is known, the latter is a procedure by means of which one is able to transport a non-scalar quantity---say, a tangent vector or a spinor or any other vector- or tensor-like object that one can think of---from one space-time point to another. Such transportation is needed to calculate the derivative of non-scalar-valued fields. The simplest example of transport rules are those for ordinary tangent four-vectors. Since it is always assumed that parallel transport is a linear operation, the transport rules in this case can be expressed with a set of 64 scalar quantities called connection coefficients and typically denoted as $\Gamma^{\alpha}_{\; \beta \mu}$. The latter have the following meaning: An arbitrary four-vector with components $V^{\alpha}$ at a point with coordinates $x^{\mu}$, when transported to a nearby point with coordinated $x^{\mu} + dx^{\mu}$, will have the components equal to 
\begin{displaymath}
V^{\alpha} - \Gamma^{\alpha}_{\; \beta \mu} V^{\beta} dx^{\mu}
\end{displaymath}
plus quantities of higher order in $dx^{\mu}$. Knowing the connection coefficients, one can evaluate the covariant derivative of an arbitrary four-vector field $U^{\alpha}$: The latter will have the components
\begin{equation}
U^{\alpha}_{\; ; \mu} = \partial_{\mu} U^{\alpha} + \Gamma^{\alpha}_{\beta \mu} U^{\beta}. 
\end{equation}

In physics, the rules of parallel transport for four-vectors are regarded as a special kind of geometry possessed by space-time, which is related to the Riemannian geometry fixed by the metric, but is not necessary determined by the latter completely. The situation is that the metric, by itself, determines certain transport rules for four-vectors; it is these rules that one considers in General Relativity. However, at a given Riemannian geometry, the actual parallel transport rules---the ones that are used to calculate the derivatives of matter fields in field equations---may differ from the rules defined by the metric. The latter fact will manifest itself in that transported four-vectors will experience addition rotation compared to the same vectors but transported according to the rules fixed by the metric. This additional rotation can be described with the quantities $S^{\alpha}_{\beta \mu}$ equal to the difference between the connection coefficients corresponding to the transport fixed by the metric and the connection coefficients corresponding to the actual transport rules. With transformation of the four-vector basis, the quantities $S^{\alpha}_{\beta \mu}$ transform as components of a four-tensor, which is called {\it torsion}, or more precisely, the {\it contorsion} tensor.

Another important case of parallel transport is that corresponding to abstract vectors and tensors associated with internal symmetry groups in elementary particle physics. Such vectors and tensors are not related in their nature to the space-time manifold and in the following will be generally referred to as {\it nonspacetime} vectors and tensors. As in the case of tangent four-vectors, one can introduce the corresponding connection coefficients, which in this case are called gauge fields. This will enable one to calculate the covariant derivative of any field whose values are nonspacetime vectors or tensors of a given kind. If $G^{i}_{j \mu}$ are the gauge fields that correspond to the considered type of nonspacetime vectors, then the components of the covariant derivative of any field $W^{i}$ whose values are such nonspacetime vectors are given by a formula similar to formula (1):
\begin{displaymath}
W^{i}_{\; ; \mu} = \partial_{\mu} W^{i} + e G^{i}_{j \mu} W^{j}, 
\end{displaymath}
where, as is customary, we have written out explicitly the `charge' $e$ of the field $W^{i}$ relative to the interaction mediated by the gauge fields $G^{i}_{j \mu}$.

It is practically always supposed implicitly that parallel transport of nonspacetime vectors is independent of torsion. The concept of bivector gauge fields offers one a scheme where this is not so. To see how it works, one should introduce the notion of the {\it bivector} derivative. Let us first define the latter for four-vector fields. For simplicity, let us consider flat space-time and introduce in it an arbitrary system of Lorentz coordinates $x^{\mu}$. The {\it covariant} derivative of an arbitrary four-vector field $\bf U$ corresponding to the transport rules fixed by the metric can then be evaluated by using the following four-step procedure: One performs an infinitesimal active {\it translation} of the differentiated field in the direction specified by the argument of the derivative; subtracts from the result the original field; divides the difference by the translation parameter $s$; and takes the limit $s \rightarrow 0$. Consider now a more general procedure where one takes the same field $\bf U$, but this time performs an {\it arbitrary} active Poincare transformation, which may include both translations {\it and} rotations. The rest of the procedure is the same as before: One subtracts from the result the original field; divides the difference by the transformation parameter $s$; and takes the limit $s \rightarrow 0$. As a result, one obtains a field which will be called the {\it bivector derivative} of field $\bf U$ and will be denoted as ${\sf D}_{\cal A} {\bf U}$, where $\cal A$ symbolizes the argument of derivative $\sf D$. Let us determine what kind of quantity this argument is.

In terms of four-vectors, an infinitesimal Poincare transformation can be described with a pair of quantities $({\bf \delta T, \delta \Omega})$, where $\bf \delta T$ is a four-vector that describes the infinitesimal translation and $\bf \delta \Omega$ is an antisymmetric four-tensor of rank two that describes the infinitesimal four-dimensional rotation. As is known, such a description of Poincare transformation is {\it not} coordinate-independent since the four-vector $\bf \delta T$ depends on the choice of the coordinate origin. At the same time, the active Poincare transformation itself is an operation that exists irregardless of the choice of the coordinate system, so it may be expected that a coordinate-independent description for it does exist. This is where five-dimensional tangent vectors and the tensors constructed out of them come in handy. It turns out that an infinitesimal Poincare transformation can be described in a coordinate-independent way with an antisymmetric {\it five}-tensor $\delta {\cal R}$ of rank two. The latter has the following components:
\begin{displaymath}
(\delta {\cal R})^{\mu \nu} =  (\delta {\bf \Omega})^{\mu \nu}, \; \;
(\delta {\cal R})^{\mu 5} = - \, (\delta {\cal R})^{5 \mu} = (\delta {\bf T})^{\mu},
\end{displaymath}
where $\mu$ and $\nu$, as all Greek indices, run  0, 1, 2, and 3, and the index value 5 corresponds to the additional fifth component (for more details the reader is referred to ref.[1]). Dividing $\delta \cal R$ by $s$ and taking the limit $s \rightarrow 0$, one obtains a finite `five-dimensional' bivector $\cal A$, which uniquely determines the corresponding infinitesimal Poincare transformation to the first order, and therefore can serve as the argument of derivative $\sf D$. This explains why the latter has been named the {\it bivector} derivative.

From the definition of the bivector derivative presented above, it is easy to obtain the following formulae for the components of $\sf D$ of an arbitrary four-vector field $\bf U$ in a Lorentz four-vector basis in flat space-time:
\begin{flushright}
\hspace*{5ex} $( {\sf  D}_{\mu 5} {\bf U})^{\alpha} =  - \, ( {\sf  D}_{5 \mu} {\bf U})^{\alpha} =\partial_{\mu} U^{\alpha}$ \hfill \hspace*{1ex} \\
\hspace*{5ex}  $( {\sf  D}_{\mu \nu} {\bf U})^{\alpha} = - \, ( {\sf  D}_{\nu \mu} {\bf U})^{\alpha} = (M_{\mu \nu})^{\alpha}_{\; \beta} \, U^{\beta}$, \hfill \rule{0ex}{3ex}
\end{flushright}
where ${\sf  D}_{\mu 5}$ denotes the bivector derivative that corresponds to translation along the axis $x^{\mu}$, ${\sf  D}_{\mu \nu}$ denotes the bivector derivative corresponding to rotation in the plane $x^{\mu} x^{\nu}$, and $(M_{\mu \nu})^{\alpha}_{\; \beta} \equiv  \delta^{\alpha}_{\, \nu} \, g_{\mu \beta} - \delta^{\alpha}_{\, \mu} \,
g_{\nu \beta}$. These formulae will also be valid in local Lorentz coordinates in space-time with arbitrary curvature. Comparing them with the formula for the components of the covariant derivative of field $\bf U$ in (local) Lorentz coordinates in space-time with arbitrary torsion:
\begin{flushleft}
\hspace*{1ex} $( \nabla_{\mu} {\bf U})^{\alpha} = \partial_{\mu} U^{\alpha} + \Gamma^{\alpha}_{\beta \mu} U^{\beta} =  \partial_{\mu} U^{\alpha} - S^{\alpha}_{\beta \mu} U^{\beta}$ \rule{0ex}{3ex}  \\ 
\hspace*{10ex} $= \partial_{\mu} U^{\alpha} - \, S^{\sigma \tau}_{\; \; \; \mu} \, \delta^{\alpha}_{\, \sigma} g_{\tau \beta} \, U^{\beta}$ \rule{0ex}{3ex} \\ \hspace*{10ex} $= \partial_{\mu} U^{\alpha} + \frac{1}{2} S^{\sigma \tau}_{\; \; \; \mu} (M_{\sigma \tau})^{\alpha}_{\, \beta} U^{\beta}$, \rule{0ex}{3ex}
\end{flushleft} 
one observes that derivatives $\sf D$ and $\nabla$ of field $\bf U$ are related to each other in the following way:
\begin{equation}
\nabla_{\mu} {\bf U} = {\sf  D}_{\mu 5} {\bf U} + \mbox{$\frac{1}{2}$} \, S^{\sigma \tau}_{\; \; \; \mu} \, {\sf  D}_{\sigma \tau} {\bf U}.
\end{equation} 
In the case of four-vector fields, the latter equation is a {\it consequence} of the definition of derivatives $\sf D$ and $\nabla$. To define the bivector derivative for a field whose values are nonspacetime vectors or tensors of some particular kind, one {\it postulates} that ({\it a}) for such a field the derivative $\sf D$ exists and that ({\it b}) it is related to derivative $\nabla$ as in equation (2).

As in the case of the covariant derivative, one can introduce the analogs of connection coefficients for $\sf D$. For the components of the bivector derivative of an arbitrary field $W^{i}$ one will then have:
\begin{equation} \begin{array}{l}
({\sf D}_{\mu 5} W)^{i} = \partial_{\mu} W^{i} + e C^{i}_{j \mu 5} W^{j} \\
({\sf D}_{\mu \nu} W)^{i} = e C^{i}_{j \mu \nu} W^{j}. \rule{0ex}{3ex}
\end{array} \end{equation}
The fields $C^{i}_{j \mu 5}$ and $C^{i}_{j \mu \nu}$ are called {\it bivector gauge fields}. With transformation in the space of corresponding nonspacetime vectors, say, $W^{i} \mapsto W'^{\, i} = L^{i}_{\, j} W^{j}$, the bivectors gauge fields transform as follows:
\begin{equation} \begin{array}{rcl}
\hspace*{-2ex} C'^{\, i}_{\; \, j \mu 5} & = & L^{i}_{\, k} \, C^{k}_{\; l \mu 5} (L^{-1})^{l}_{\, j} + e^{-1} L^{i}_{\, k} \partial_{\mu} (L^{-1})^{k}_{\, j} \\ 
\hspace*{-2ex} C'^{\, i}_{\; \, j \mu \nu} & = & L^{i}_{\, k} \, C^{k}_{\; l \mu \nu} (L^{-1})^{l}_{\, j}. \rule{0ex}{3ex}
\end{array} \end{equation}
So the fields $C^{i}_{j \mu 5}$ transform as ordinary gauge fields, while the fields $C^{i}_{j \mu \nu}$ transform as components of a tensor of rank (1,1) over the space of corresponding nonspacetime vectors. 

If one assumes that the Lagrangian density for matter fields in local field theory depends on the value of the field at a given point and on the value of its {\it covariant} derivative, then the considered scheme implies that in the absence of torsion the matter fields are affected only by the gauge fields $C^{i}_{j \mu 5}$, and it is the latter that play the role of `ordinary' gauge fields in this case: $G^{i}_{j \mu} = C^{i}_{j \mu 5}$. Under such circumstances, the fields $C^{i}_{j \mu \nu}$ have no direct effect on matter fields irregardless of whether they are nonzero or not. At nonzero torsion, matter fields `feel' both types of bivector gauge fields, and one has
\begin{equation}
\mbox{$G^{i}_{j \mu} = C^{i}_{j \mu 5} + \frac{1}{2} C^{i}_{j \sigma \tau} S^{\sigma \tau}_{\; \; \; \mu}.$}
\end{equation} 
Figuratively speaking, the fields $C^{i}_{j \mu \nu}$ {\it translate} the additional rotation of tangent four-vectors which torsion is, into additional rotation of the relevant nonspacetime vectors. One can also see that the affine geometry represented by the gauge fields $C^{i}_{j \mu \nu}$ is, so to say, {\it hidden} in the sense that it has a noticeable direct effect on matter fields only when torsion is sufficiently large. Fortunately, as it will be shown below, the existence of this type of bivector gauge fields should also manifest itself {\it indirectly}: in that the fields $C^{i}_{j \mu 5}$---whose effect on matter fields is not suppressed by torsion, and which have all the properties of conventional gauge fields---{\it created} by matter fields will {\it differ} from the gauge fields the same matter fields would have created within the traditional gauge field theory framework.

\vspace{1ex}

The rest of this paper is devoted to the bivector gauge fields associated with electromagnetism. In this case the fields  $C_{\mu 5}$ and $C_{\mu \nu}$ will apparently have only two lower space-time indices and no indices associated with internal degrees of freedom. In order to construct the Lagrangian density that would determine the dynamics of such fields, one should first introduce the analog of the field strength tensor for the derivative $\sf D$. As is shown in part VI of ref.[1], the latter is a `five-dimensional' tensor ${\sf F}$ whose components have four lower indices and the following symmetry properties:
\begin{displaymath} \begin{array}{l}
{\sf F}_{ABCD} = - \, {\sf F}_{BACD} = - \, {\sf F}_{ABDC} \\ 
{\sf F}_{ABCD} = - \, {\sf F}_{CDAB}, \rule{0ex}{3ex}
\end{array} \end{displaymath}
where indices $A$, $B$, $C$, and $D$ run  0, 1, 2, 3, and 5. The components of  ${\sf F}$ are  the following:
\begin{displaymath} \begin{array}{l}
{\sf F}_{\mu 5 \alpha 5} = iF_{\mu \alpha}, \; {\sf F}_{\mu 5 \alpha
\beta} = - {\sf F}_{\alpha \beta \mu 5} = \partial_{\mu} C_{\alpha \beta},
\\
{\sf F}_{\mu \nu \alpha \beta} = - \; g_{\mu \alpha} C_{\nu \beta} +
g_{\nu \alpha} C_{\mu \beta} \rule{0ex}{3ex} \\ 
\hspace{18ex} + \; g_{\mu \beta} C_{\nu \alpha} - g_{\nu
\beta} C_{\mu \alpha}, \rule{0ex}{3ex}
\end{array} \end{displaymath}
where we have introduced the notation $A_{\alpha} = -iC_{\alpha 5}$ and where $F_{\alpha \beta} = \partial_{\alpha} A_{\beta} - \partial_{\beta} A_{\alpha}$. One may then suppose that, similar to the case of ordinary gauge fields, the Lagrangian density for their bivector analogs should be some bilinear combination of the components of ${\sf F}$. Observing now that from the latter one can construct only {\it two} independent true scalars, for example,
\begin{displaymath}
{\sf F}^{ABCD} {\sf F}_{ABCD} \; \; \mbox{and} \; \; {\sf F}^{AC}_{\hspace{2ex} AD} \, {\sf F}^{BD}_{\hspace{2ex} BC}, 
\end{displaymath}
one obtains the following general expression for the Lagrangian density:
\begin{equation}
a \cdot {\sf F}^{ABCD} {\sf F}_{ABCD} + b \cdot {\sf F}^{AC}_{\hspace{2ex}AD}
\, {\sf F}^{BD}_{\hspace{2ex} BC} \; ,
\end{equation}
where $a$ and $b$ are some unknown coefficients. Determining the value of the latter from the requirement that combination (6) reproduce the standard kinetic term for the field $A_{\alpha}$ (which can be regarded as ordinary electromagnetic potential) and the kinetic term for the antisymmetric tensor field (see e.g.\ ref.[3]), one arrives at the following Lagrangian density:
\begin{equation} \begin{array}{l}
{-\scriptstyle \frac{1}{4}} F^{\alpha \beta} \! F_{\alpha \beta} +
{\scriptstyle \frac{1}{4}} \, \partial^{\mu} K^{\alpha \beta} \partial_{\mu}
K_{\alpha \beta} - (\partial^{\mu} K_{\mu \alpha})^{2} \\ \hspace{14ex} 
+ \; 2 \kappa \, F^{\alpha \beta} \! K_{\alpha \beta} - {\scriptstyle \frac{3}{2}}
\kappa^{2} \, K^{\alpha \beta} \! K_{\alpha \beta} \rule{0ex}{3ex}.
\end{array} \end{equation}
where $K_{\alpha \beta} = -i \kappa C_{\alpha \beta}$. Here $\kappa$ is a certain universal constant with dimension of inverse length that appears within the theory of five-dimensional tangent vectors, and whose value is not fixed by mathematics (see ref.[1]). As one can see, in addition to the kinetic terms for $A_{\alpha}$ and $K_{\alpha \beta}$, one obtains a term where $A_{\alpha}$ and $K_{\alpha \beta}$ mix and a term that has the form of a mass term for the field $K_{\alpha \beta}$. Owing to the first of these additional  terms, the equations for the electromagnetic field and for the field $K_{\alpha \beta}$ are no longer independent from each other and have the following form:
\begin{equation}
\partial^{\alpha} \! F_{\alpha \beta} = 4 \kappa \, \partial^{\alpha}
\! K_{\alpha \beta} + j_{\beta} ,
\end{equation}
\begin{equation} \begin{array}{c}
\partial^{2} \! K_{\alpha \beta} +
2 \, \partial^{\lambda} ( \partial_{\alpha} \! K_{\beta \lambda} -
\partial_{\beta} \! K_{\alpha \lambda} )  + 6 \kappa^{2} K_{\alpha \beta} \\ 
= 4 \kappa F_{\alpha \beta} + j_{\alpha \beta} , \rule{0ex}{3ex}
\end{array} \end{equation}
where $j_{\alpha}$ and $j_{\alpha \beta}$ are obtained by varying the part of the action that describes the interaction of matter fields with bivector gauge fields with respect to $A_{\alpha}$ and $K_{\alpha \beta}$, respectively. It is useful to observe that under the assumption made above about Lagrangian density depending on the {\it covariant} derivative of matter fields, one has
\begin{displaymath}
j_{\alpha \beta} = \kappa^{-1} S_{\alpha \beta}^{\; \; \; \; \mu} \, j_{\mu},
\end{displaymath}
where $S_{\alpha \beta}^{\; \; \; \; \mu} = g_{\alpha \sigma} \, g_{\beta \tau} \, S^{\sigma \tau}_{\; \; \; \; \omega} \, g^{\omega \mu}$. So the components $j_{\alpha \beta}$ will be comparable with $j_{\alpha}$ only if contorsion is of the order of constant $\kappa$.\footnote[3]{Another possibility is that for some matter fields the Lagrangian density would depend on their bivector derivative {\it directly}, not via $\nabla$ (see e.g.\ ref.[4]).}

\vspace{1ex}

As it has been mentioned above, apart from its direct action on matter fields (which is suppressed by torsion), the field $K_{\alpha \beta}$ manifests its existence in that the electromagnetic potential $A_{\alpha}$ created by some set of sources will {\it differ} from the potential these same sources would have created according to conventional electrodynamics. To see that this is indeed so, let us derive from the general equations (8) and (9) the equations that specifically determine the electromagnetic potential. To this end, let us calculate the 4-divergence of both sides of equation (9) with respect to index $\alpha$. Multiplying both sides of the equation by $\frac{2}{5} \kappa^{-1}$ and introducing the notation $C_{\beta} \equiv \frac{2}{5} \kappa^{-1} \partial^{\alpha} K_{\alpha \beta}$, one obtains
\begin{center}
\mbox{$- \partial^{2} C_{\beta} + 6 \kappa^{2} C_{\beta} =  \frac{8}{5} \, \partial^{\alpha} \! F_{\alpha \beta} + \frac{2}{5} \kappa^{-1} \partial^{\alpha} \! j_{\alpha \beta} .$}
\end{center}
Substituting the right-hand side of equation (8) instead of $\partial^{\alpha} \! F_{\alpha \beta}$ and rearranging the terms, one gets
\begin{equation}
\mbox{$\partial^{2} C_{\beta} + 10 \kappa^{2} C_{\beta} = -  \frac{8}{5} \, t_{\beta} - \frac{8}{5} \, j_{\beta},$}
\end{equation}
where $t_{\beta} \equiv \frac{1}{4} \kappa^{-1} \partial^{\alpha} \! j_{\alpha \beta}$. If the electromagnetic potential obeys the Lorentz condition $\partial^{\alpha} \! \! A_{\alpha} = 0$, equation (8) acquires the form
\begin{equation}
\partial^{2} \! A_{\beta} = 10 \kappa^{2} C_{\beta} + j_{\beta}.
\end{equation}
Expressing $10 \kappa^{2} C_{\beta}$ in terms of $j_{\beta}$, $t_{\beta}$, and $\partial^{2} C_{\beta}$ via equation (10), one can present equation (11) as
\begin{center}
\mbox{$\partial^{2} (A_{\beta}+ C_{\beta}) =(1 -  \frac{8}{5}) j_{\beta} - \frac{8}{5} t_{\beta}.$}
\end{center}
Finally, introducing the notation $B_{\beta} \equiv A_{\beta}+ C_{\beta}$, one arrives at the following equations:
\begin{center} 
\hfill \hspace*{9.5ex} \mbox{$\partial^{2} B_{\beta} = - \frac{8}{5} t_{\beta} -  \frac{8}{5} j_{\beta} + j_{\beta}$} \hfill {\rm (12a)} \\ \hfill \mbox{$10 \kappa^{2} C_{\beta} + \partial^{2} C_{\beta} = -  \frac{8}{5} \, t_{\beta} - \frac{8}{5} \, j_{\beta}$}  \rule{0ex}{3ex} \hfill  \hspace*{2.5ex} {\rm (12b)} \\  \hfill  \hspace*{5.5ex} \mbox{$A_{\beta} = B_{\beta} - C_{\beta}$}. \rule{0ex}{3ex} \hfill {\rm (12c)}
\end{center} \setcounter{equation}{12}

As one can see, if the field $C_{\beta}$ were massless, too, the contributions to $B_{\beta}$ and $C_{\beta}$ created by the source $-\frac{8}{5} t_{\beta} - \frac{8}{5} j_{\beta}$ would be exactly the same and would cancel out each other in the potential $A_{\beta}$. So the latter would be exactly what it should be according to conventional electrodynamics. To see how the nonzero mass of $C_{\beta}$ changes the potential, let us consider the solution of equations (12) for a static point source at the coordinate origin. As usual, `static' and `point' mean that $j_{i} = 0$ and $j_{0} = e \delta(\vec{r})$ and that all the time derivatives are zero. For simplicity, we will also suppose that torsion is so small that the contribution of $t_{\alpha}$ to the electromagnetic potential is insignificant. The solution is the following:
\begin{displaymath}
\mbox{$B_{i} = C_{i} = 0, \; \; B_{0} = (1 - \frac{8}{5}) \, \frac{e}{4 \pi r}, \; \; C_{0} = - \frac{8}{5} \, \frac{e}{4 \pi r} e^{-\mu r} ,$}
\end{displaymath}
where $\mu = \sqrt{10} \kappa$. Thus, for the electromagnetic potential one has:
\begin{equation}
\mbox{$A_{i} = 0$ and $A_{0} = \frac{e}{4 \pi r} - \frac{8}{5} \, \frac{e}{4 \pi r} (1 - e^{-\mu r}).$}
\end{equation}
It is apparent that at $r \ll \mu^{-1}$, the second term in the expression for $A_{0}$ is practically zero, and the scalar potential itself is given by the usual formula: $A_{0} = (e / 4 \pi r)$. As $r$ increases, the second term in the expression for $A_{0}$ grows, and at $r =  {\rm ln} (\frac{8}{3}) \, \mu^{-1} = 0.9808 \, \mu^{-1}$ the potential turns to zero and then changes its sign, as is shown in Figure 1. 
\begin{figure}[h!]
\centering
\includegraphics[width=0.45\textwidth]{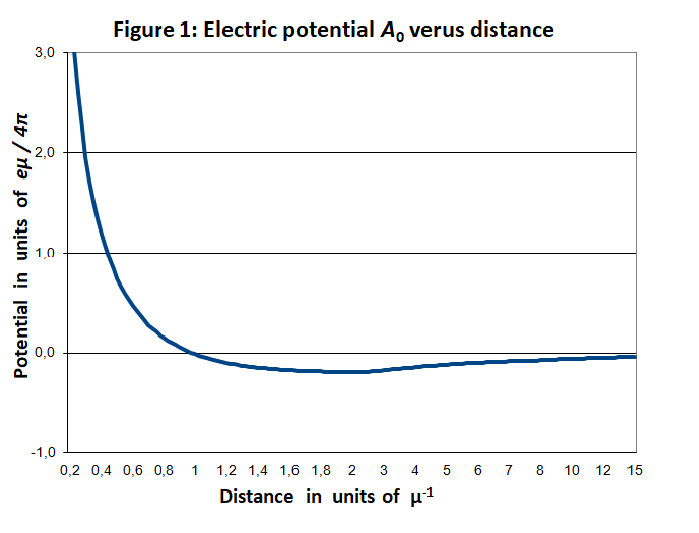}
\end{figure}
Here the potential is measured in the units of $e \mu / 4 \pi$ and $r$  is measured in the units of $\mu^{-1}$. With further increase of $r$, one observes a `valley', where the potential changes very little and after reaching a minimum, begins to grow again. This means that within this distance range, the electric field decreases and turns to zero and then changes its direction for the opposite. At $r \gg \mu^{-1}$, one has  $A_{0} = - \frac{3}{5} \cdot ( e / 4 \pi r)$, as if this was a Coulomb potential created by the charge $- \frac{3}{5} e$. 
As in conventional electrodynamics, the corresponding electric field will have only the radial component, which can be presented in the following form:
\begin{displaymath}
\mbox{$E_{r} =  \frac{e \mu^{2}}{4 \pi} {\cal E} (\mu r)$},
\end{displaymath}
where
\begin{displaymath}
\mbox{${\cal E} (z) = \frac{8}{5} \frac{ e^{-z}}{\rule{0ex}{1.5ex} z^{2}} \, (1+z) - \frac{3}{5} \frac{1}{\rule{0ex}{1.5ex} z^{2}}.$}
\end{displaymath}
As one can see, at distances much less than $\mu^{-1}$, the interaction between two charged particles obeys the Coulomb law. The behavior of the electric field at $r$ greater than $\mu^{-1}$ is shown in Figure 2. 
\begin{figure}[h!]
\centering
\includegraphics[width=0.45\textwidth]{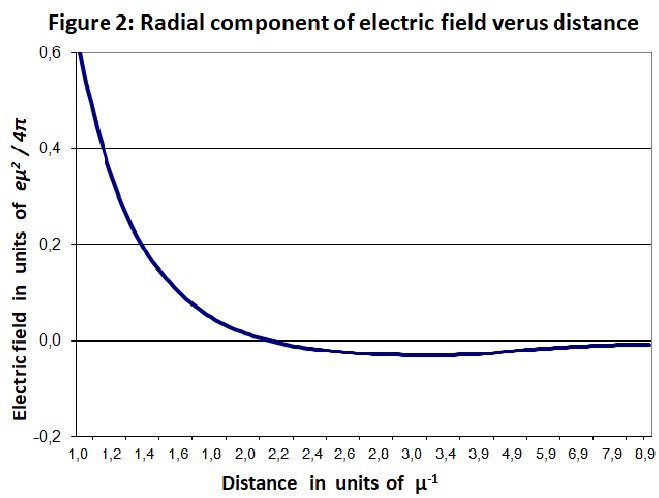}
\end{figure}
As the distance between the particles increases and approaches a critical value of $r_{c} = 2.118 \, \mu^{-1}$, the magnitude of the force acting between them rapidly diminishes and turns to zero at $r = r_{c}$. With further increase of the distance, the electrostatic interaction between the particles appears again, but now it is of opposite direction: charges of the same sign now {\it attract} and charges of opposite signs {\it repulse}. With further increase of $r$, the interaction between the particles continues to grow until it reaches a maximum around $r = 3.224 \, \mu^{-1}$, after which its magnitude begins to decrease. At $r \gg \mu^{-1}$, the force of interaction between the particles is again inversely proportional to $r^{2}$, only now its magnitude is smaller than that of Coulomb interaction in conventional electrostatics by a factor of $\frac{3}{5}$, and one has attraction instead of repulsion and vice versa. 

\vspace{1ex}

From all these observations one can conclude that constant $\kappa$ should be either very small or very large, so that the deviation of potential (13) from Coulomb form would lie outside the range where the Coulomb law has been tested. If $\kappa$ were {\it large}, and accordingly, the distance $\mu^{-1}$ were small, the described unorthodox behavior of the electric field would be a {\it quantum} effect and should be examined within quantum field theory. However, at macroscopic distances one should then have observed attraction of charges of the same sign and repulsion of charges of opposite signs, which is not what one has in reality.

One is therefore left with the possibility that constant $\kappa$ is {\it small}, and accordingly, the distance $\mu^{-1}$ is large. In this case, the considered deviation from the Coulomb law is a {\it macroscopic} effect and is actually a deviation from classical electrodynamics. It is obvious that measuring the deviation from the Coulomb law at large distances directly is not an option. A more realistic way to confirm the existence of bivector gauge fields, and to measure the constant $\kappa$, is to examine monochromatic radiation from far away sources. To simplify the calculations, let us again suppose that the contribution of $t_{\beta}$ is insignificant. Then, according to equations (12), at large distances from the source, the radiation would approximately be a superposition of two monochromatic plane waves: one with the wavelength $\lambda_{1} = 2 \pi c \omega^{-1}$ and relative amplitude $-\frac{3}{5} = 1 - \frac{8}{5}$, the other with the wavelength $\lambda_{2} = 2 \pi (\omega^{2} c^{-2} - \mu^{2})^{-1/2}$ and relative amplitude $+\frac{8}{5}$, where $\omega$ is the common angular frequency of both waves. Let us suppose that the source is located at the coordinate origin. If $\mu^{-1}$ is a very large macroscopic distance, then for all the electromagnetic radiation recorded from the source one evidently has $\mu c / \omega \ll 1$. This means that one can safely disregard the longitudinal component of the wave with the wavelength $\lambda_{2}$ and that, with the same precision, the transversal components of the total electromagnetic field will be proportional to
\begin{equation}
\mbox{$- \frac{3}{5} \, cos ( \omega t - 2 \pi x / \lambda_{1} ) + \frac{8}{5} \, cos ( \omega t - 2 \pi x / \lambda_{2} ),$}
\end{equation}
where, for simplicity, it is assumed that the waves have linear polarization and propagate along the $x$ axis and that zero time has been chosen appropriately. Let us examine expression (14) at fixed $t$ and different $x$. Observing that
\begin{displaymath}
\frac{2 \pi}{\lambda_{2}} \approx \frac{2 \pi}{\lambda_{1}} (1 - \frac{\mu^{2} c^{2}}{2 \omega^{2}}) = \frac{2 \pi}{\lambda_{1}} - \frac{\mu^{2} c}{2 \omega},
\end{displaymath}
one obtains
\begin{displaymath}
\mbox{$- \frac{3}{5} cos (  \omega t - 2 \pi x / \lambda_{1} ) + \frac{8}{5} cos (  \omega t - 2 \pi x / \lambda_{1} +  \mu^{2} c x / 2 \omega ).$}
\end{displaymath}
By using standard trigonometric formulae, one can cast the above expression into the following form:
\begin{displaymath}
\xi \cdot cos ( \omega t - 2 \pi x / \lambda_{1} + \phi_{0} ) , 
\end{displaymath}
where
\begin{displaymath}
\mbox{$ \xi \equiv  \frac{8}{5} \sqrt{ ( \, cos ( \mu^{2} c x / 2 \omega ) - \frac{3}{8} \,  )^{2} + sin^{2} ( \mu^{2} c x / 2 \omega ) }$} 
\end{displaymath}
and
\begin{displaymath} \begin{array}{l}
\hspace*{4ex} \mbox{$ cos \, \phi_{0} \equiv  \frac{8}{5} \, ( \, cos ( \mu^{2} c x / 2 \omega )  - \frac{3}{8} \, ) \cdot \xi^{-1} $}  \\ 
\hspace*{4ex} \mbox{$sin \, \phi_{0} \equiv \frac{8}{5} \, sin ( \mu^{2} c x / 2 \omega ) \cdot \xi^{-1} .$} \rule{0ex}{4ex}
\end{array} \end{displaymath}
Since the size of the measuring equipment is sure to be much smaller than the distance to the source, $\phi_{0}$ will be practically a constant phase shift and therefore can be neglected. One can also see that the {\it intensity} of the radiation considered will differ by the factor $\xi^{2}$ from the intensity of the radiation the same source should have emitted according to conventional electrodynamics. At $x=0$ this factor is unity, as it should be. As $x$ increases, $\xi^{2}$ monotonically grows until it reaches its maximum of $(2.2)^{2} = 4.84$ at $x=2 \pi \omega / \mu^{2} c \equiv x_{c}$. After that, $\xi^{2}$ monotonically decreases to its initial unity value, which it reaches at $x=2 x_{c}$. The behavior of the factor $\xi^{2}$ versus $x/x_{c}$ becomes obvious if one observes that
\begin{equation}
\xi^{2} = 2.92 - 1.92 \cdot cos ( \pi x / x_{c} ).
\end{equation}
It is shown in Figure 3. 
\begin{figure}[h!]
\centering
\includegraphics[width=0.45\textwidth]{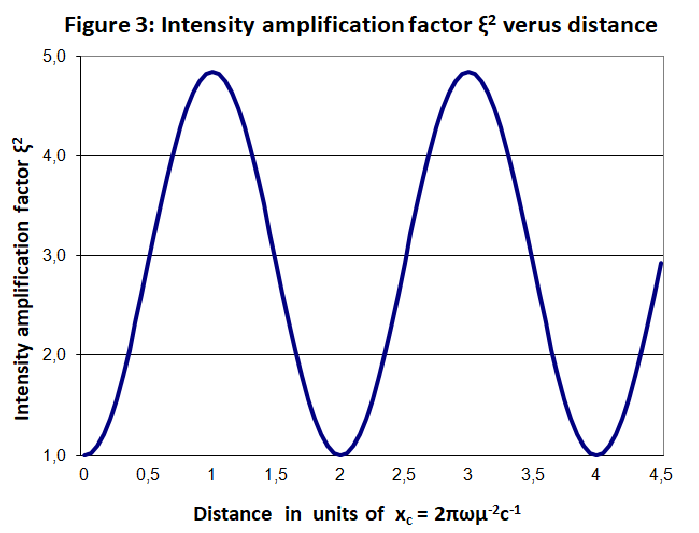}
\end{figure}
One should notice that the distance $x_{c}$ depends {\it quadratically} on the length parameter $\mu^{-1}$. It is also inversely proportional to the radiation wavelength: $x_{c} = 4 \pi^{2} / \mu^{2} \lambda_{1}$. The latter fact implies that for a given (unknown) $\mu$, the value of $x_{c}$ for radiowaves is smaller than it is for infrared radiation; the value of $x_{c}$ for infrared radiation is smaller than it is for visible light; and so on. In other words, the larger the wavelength, the bigger value of the length parameter $\mu^{-1}$ one is able to measure. For example, if the largest distance one can make measurements at is of the order of 10 billion light years, then for infrared radiation with wavelength $\sim 10^{-4}$ cm, the effect can be observed if $\mu^{-1} \leq 10^{11}$ cm, whereas for radiowaves with wavelength $\sim 1$ m, the effect will be observable if  $\mu^{-1} \leq 10^{14}$ cm.

If one could record radiation coming from outer space with {\it any} wavelength, the considered intensity boosting effect would be observed for sure if only the bivector gauge fields associated with electromagnetism do exist and if their dynamics is indeed described by Lagrangian density (7). In reality, there are obvious limitations to one's ability to record low frequencies. To see what this means in relation to observing the effect described, let us rewrite formula (15) in the following way:
\begin{equation}
\xi^{2} = 2.92 - 1.92 \cdot cos ( \pi \omega_{c} / \omega ),
\end{equation}
where $\omega_{c} = \mu^{2} cx / 2 \pi$ and $x$ is the distance to the source. The behavior of $\xi^{2}$ versus $\omega / \omega_{c}$ is shown in Figure 4.
\begin{figure}[h!]
\centering
\includegraphics[width=0.45\textwidth]{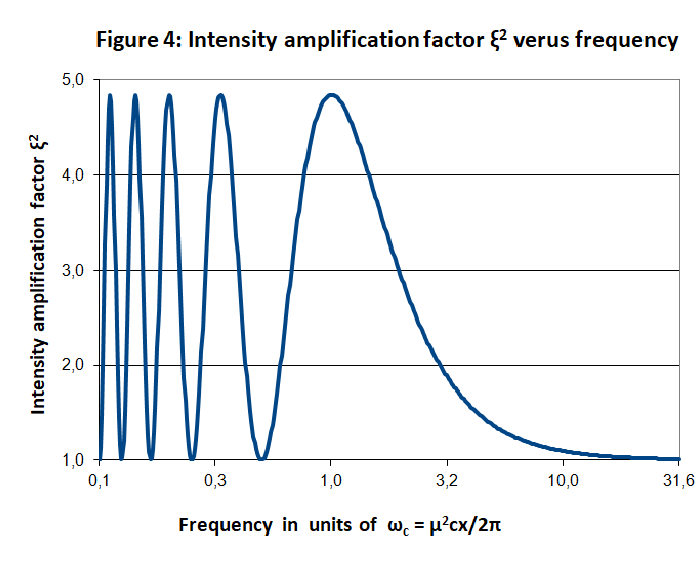}
\end{figure}
As one can see, the effect could be observed only if the lowest frequency one is able to record is of the order of or smaller than $\omega_{c}$. 

It should be emphasized that to be able to attribute the increase in intensity of radiation emitted by distant cosmic sources to the existence of bivector gauge fields, one has to observe the characteristic patterns both in the variation of intensity with distance to the source for radiation with same frequency (as is depicted in Figure 3) and in the variation of intensity with frequency for radiation coming from a single source (as is shown in Figure 4).

\end{document}